\documentclass{article}

\usepackage[dvips]{graphicx}

\begin{document}

\title{
Free energy of multicomponent embryo formation
}

\author{Victor Kurasov}

\date{Victor.Kurasov@pobox.spbu.ru}

\maketitle

This publication continues description presented in
\cite{Him}
taking into account all definitions and facts from
\cite{Him}.

Due to the compact character of the embryo of the liquid phase one can
consider it in the state of the internal equilibrium.

In the capillarity approximation the free energy of the embryo consists
of the volume part and of the surface part. The surface part $F_S$
has the form  $$F_S = S\gamma\ \ , $$  where  $\gamma$
is the  surface tension,
$S$
is the  square of the  surface.
In the capillarity approximation it is supposed that the surface tension
is the surface tension of the plane surface.

The volume part  $F_V$
of the free energy $F$
has the form
          $$ F_V = - \sum_i b_i \nu_i  \ \ , $$
where $\nu_i$
are the numbers of the molecules of the different components inside the
embryo,
$b_i$
are the excesses of the chemical potentials in the bulk phase calculated
from the saturated case,
index  $i$
has the values
$1,..,m$
for the components in the embryo.
The set $\{\nu_i\}$
of the numbers of the different components inside the embryo is the most
simple set from the kinetic point of view as far as the numbers of the
molecules
varies separately in the elementary act of interaction.

The bulk phase is considered in the frames of the capillarity approximation
as the homogeneous solution.

For  $F$
one can get
\begin{equation}
F = - \sum_i  \nu_i b_i  + S \gamma
\end{equation}

{\bf
Gibbs-Duhem equations }

For  $b_i$
one can come in frames of the standard theory of solutions to the following
equations
\begin{equation}
b_i = \ln(\frac{n_i}{n_{i\infty}(\xi)}) \ ,
\end{equation}
where there is the
molecule number density of the saturated vapor of the given concentration
 in the denominator.
The concentrations $\xi_i$
of the  component $i$
form the vector $\xi$.
They satisfies to $\sum_i \xi_i = 1$.
In the further considerations we shall use the definition $\xi$
for $m$ dimensional vector or for $m-1$ dimensional vector without any
special notations.

According to the thermodynamic definitions of the activity coefficients
$f_i$
\begin{equation}
n_{i \infty} (\xi) \equiv n_{i \infty} |_{\xi_i=1}  \xi_i f_{i}(\xi)
\ \ .
\end{equation}
The parts of the chemical potential  which depend on these coefficients
are the regular functions in the limit situations. This property extracts
the activity coefficients.

The activity coefficients satisfy the equations analogous to the Gibbs-Duhem
one
\begin{equation}\label{3.1.1.4}
\sum_a \xi_a d \ln(f_a)  = 0
\ \
\end{equation}
simply due to the chosen function dependencies.
This is the consequence of the expression for the differential of the free
energy
$$
dF_V = - \sum_i b_i \ d \nu_i
\ \ .
$$

To fulfill the further considerations  we  have to notice the following
very important fact, which will be called as the principle of the
correspondence.
 {\bf The Gibbs-Duhem equation has to  be written in the same geometrical
construction as the approximation for the free energy.}
The problem of the construction of the Gibbs-Duhem
equation is connected with the concrete approximation in which the free
energy is written. In the capillarity approximation the surface tension
is taken as the surface tension of the plane surface. This point has the
principal significance. For the spherical embryo the square of the surface
is connected with the number of the embryo. For the plane film the square
of the surface is the independent parameter. So, in the Gibbs-Duhem equation
it has to be considered as the independent one.
So, the differential of the free energy can be written as
$$
dF = - \sum_i b_i \ d \nu_i + \gamma d S
\ \ .
$$

For the embryo as the object of the Gibbs-Duhem equation it can be written
\begin{equation}
- \sum_i \nu_i  d \ln(f_i)
+
S d \gamma = 0
\ \ .
\end{equation}

>From the first point of view the natural definition of the concentration
is the following one
\begin{equation}
\xi_i = \frac{\nu_i}{\sum_j \nu_j }
\ \ .
\end{equation}
Then
\begin{equation}\label{3.1.1.7}
\sum_i \xi_i d \ln(f_i)
 -
\tilde{S} d\gamma = 0 \ ,
\end{equation}
where
 $\tilde{S}=S/(\sum_i \nu_i)$.
As far as the first sum has to go to zero one can  get
$$ \tilde{S} d\gamma = 0 $$

The last equation is rather naive one and is gotten with  account
of the Gibbs-Duhem equation for the whole embryo which doesn't satisfy
to the mentioned principle of the
 correspondence. Certainly, it is impossible
to believe that one can not differentiate the surface tension
on the concentration.

The last equation is presented as far as it accumulates the phenomenological
rule of Wilemski and Renninger which forbid to differentiate the surface
tension in the expression for the saddle point.  The connection with the
geometrical
aspects of the approximation has not been established earlier. Now we
see that this is really the paradox and this paradox is the consequence
of the approximation for the free energy.

Note that when the microscopic corrections due to the curvature are taken
into account, the Gibbs-Duhem equation must be also corrected in the
corresponding
terms due to the changed geometry.

{\bf
The physical reason for the Wilemski-Renninger paradox
}

The physical reason of the Wilemski-Renninger paradox is existence of
the surface layer with the changed concentration. The difference of this
concentration from the bulk value compensates the variations of the surface
tension. In some naive model one can say that it is profitable for the
molecules
of some component to move to the surface layer to decrease the surface
tension. Certainly, they will move until the moment of some stability
when the further variation of the concentration will  not  lead  to
decrease
of the surface tension, i.e. when the derivative of the surface is equal
to zero. Stress again that this qualitative picture is rather naive, but
it will open the way to construct  the capillarity model with the
surface excesses which will be done later.

{\bf
The difference in the sets of the independent variables in the determination
of the saddle point and in the Gibbs-Duhem equation.  }

Here some certain disadvantages of the
mentioned model will be presented.
The Gibbs-Duhem equation was derived with account that the chemical
potentials are not connected with the number of the molecules of the components
in the embryo. But this connection exists and can be taken into account.
This account leads to the two different approaches.

Neglect this dependence, i.e. put
$(\partial \nu_i / \partial \nu_j) = \delta_{ij}$, and get
\begin{equation}
(\frac{\partial F}{\partial \nu_i}) \mid_{\nu_j \ \xi_j } =
-    b_i    +    (\frac{2\gamma}{r})
 v_{l \ i}
\ \ ,
\end{equation}
where $r$
is the radius of the embryo,
$$
       v_{l \ i} = (\frac{\partial     V}{\partial \nu_i})
$$
is the volume for one molecule in the liquid phase, $V$ is the volume
of the embryo.
Then in the saddle point
\begin{equation}
\frac{b_a(\xi)}{v_{la}}
=
\frac{b_b(\xi)}{v_{lb}}
\ \ .
\end{equation}

Express $(\partial \nu_i / \partial \nu_j)$ through the concentrations
(due to the finite number of the molecules in the embryo) and get
$$
\frac{\partial F}{\partial \nu_i} =
 -  \sum_j b_j \frac{\partial \nu_j}{\partial \nu_i} +
\gamma \frac{\partial S}{\partial \nu_i}
\ \
$$
or
$$
\frac{\partial F}{\partial \nu_i} =
 -   \frac{b}{\xi_i}  +
\gamma \frac{2}{r}\frac{v}{\xi_i}
\ \ ,
$$
where
$$
v =
\sum_j v_{l \ j} \xi_j
\ \ ,
$$
$$
b = \sum_j \xi_j  b_j
\ \ .
$$
One and the same equation appears for every component
$$
b  =
\gamma \frac{2}{r} v
\ \
$$
for the saddle point.

Also there appears the certain difficulty connected with the absence of
the
connection between the concentrations and the numbers of the molecules
in the embryo (the integral definition of the concentration
 is only some approximation). As a result one can propose
the capillarity model with the surface excesses of components

{\bf
The excesses of the components}

To give more justified value of the free energy one can use the formalism
of the Gibbs method of the dividing surfaces. The formalism prescribes to
put on  the considered surface the excesses  $\tilde{\nu_i}$ (consider
the surface of tension)
of the molecules of component $i$.

Then the concentration can be defined as
$$
\xi_i = \frac{\nu_i - \tilde{\nu_i}}
{\sum_j(\nu_j - \tilde{\nu_j})}
\ \ .
$$
This leads to
$$
- \sum_i \xi_i d b_i
+
\sum_i \frac{\tilde{\nu_i}}{\sum_k (\nu_k - \tilde{\nu_k})}
d b_i
+ \frac{S}{\sum_k (\nu_k - \tilde{\nu_k})}
d \gamma
 = 0
$$
with the concrete realization
$$
- \sum_i \xi_i \frac{d b_i}{d \xi_j}
+
\sum_i \frac{\tilde{\nu_i}}{\sum_k (\nu_k - \tilde{\nu_k})}
\frac{d b_i}{d \xi_j}
+ \frac{S}{\sum_k (\nu_k - \tilde{\nu_k})}
\frac{d \gamma}{d \xi_j}
 = 0
\ \ .
$$

Due to concrete form of the activity coefficients the first term goes
to zero and
$$
\sum_i \frac{\tilde{\nu_i}}{\sum_k (\nu_k - \tilde{\nu_k})}
d b_i
+ \frac{S}{\sum_k (\nu_k - \tilde{\nu_k})}
d \gamma
= 0
$$
with the concrete realization
$$
\sum_i \frac{\tilde{\nu_i}}{\sum_k (\nu_k - \tilde{\nu_k})}
\frac{d b_i}{d \xi_j}
+ \frac{S}{\sum_k (\nu_k - \tilde{\nu_k})}
\frac{d \gamma}{d \xi_j}
= 0
\ \ .
$$

The last relation forms the system of $m-1$ equations. Then all excesses
are defined with one arbitrary constant. This constant can be fixed if
the distance between the considered  surface and the equimolecular surface
is known.
It is natural to consider the surface of tension as the basic surface
as far as the tension is prescribed  only to this surface.

On the level of the differentiates one can get
$$
\sum_i (\nu_i - \tilde{\nu}_i)  d b_i
+ \sum \tilde{\nu_i} d b_i  -  S d\gamma = 0
\ \ .
$$
The first sum goes to zero which leads to
$$
\sum \tilde{\nu_i} d b_i  - S d\gamma = 0
$$
or
$$
\sum_i \frac{\tilde{\nu}_i}{S} d b_i = d \gamma
\ \ .
$$

{\bf
The global structure of the free energy
}

At first we shall describe the global structure without excesses of the
components of the embryo.

In terms of
$
b(\xi) $
 then
\begin{equation}
F = -\nu b(\xi) + S \gamma
\ \ ,
\end{equation}
where
$$
\nu = \sum \nu_i
\ \ .
$$

To have the condensation it is necessary and sufficient to have the positive
values of
    $b(\xi)$.
Introduce
\begin{equation}
B(\xi) = \frac{b(\xi) }{ 6 \pi^{1/2} \gamma^{3/2} v}
\ \ .
\end{equation}
Suppose that the volumes for one molecule depend on the concentration weakly.
Introduce
\begin{equation}
\kappa = (S\gamma)^{3/2} = 6 \pi^{1/2} \nu \gamma^{3/2}
v
\ \ .
\end{equation}
Then
\begin{equation}
F = -\kappa B(\xi) + \kappa^{2/3}
\ \ .
\end{equation}

All channels are the straight lines, they start at the origin of the
coordinates
of the number of the molecules in the embryo and don't cross. The independent
consideration of every channel is possible.

For the characteristics of the saddle point one can get
\begin{equation}
\kappa_m = (\frac{2}{3B(\xi)})^3
\ \ ,
\end{equation}
\begin{equation}
F_m = \kappa_m^{2/3}(\xi) / 3
\ \ .
\end{equation}

{\bf
Global structure with excesses
}

Under the macroscopic description
\begin{equation}
\tilde{\nu_a} = \lambda_a S
\ \ .
\end{equation}
Certainly,
$$
\lambda_i = \lambda_i (r)
\ \ .
$$
But the application of the Gibbs-Duhem equation for the film requires
the plane surface. Then it is necessary to put
$$
\lambda_i = \lambda_i (r = \infty)
\ \ .
$$

The mentioned property is the concrete realization of the principle of
the "unique curvature":
{\bf
All surface characteristics and the Gibbs-Duhem  equation  must  be
referred
to one and the same surface.}

Then
\begin{equation}
F = -  \sum_i  [ b_i (\xi) \xi_i + b_i (\xi) \xi_i
\frac{\tilde{\nu_i}}{\nu_i-\tilde{\nu_i}} ]
(\sum_a (\nu_a - \tilde{\nu_a} ) ) + S \gamma
\ \ .
\end{equation}

Let the surface of tension $S$
be the surface around all molecules inside the embryo (to be
the equimolecular
surface). Then
\begin{equation}
\sum_a (\nu_a - \tilde{\nu_a} ) v_{l\ a} =
\frac{S^{3/2}}{6 \pi^{1/2}}
\ \ ,
\end{equation}
\begin{equation}
F =
-
[ \sum_i \xi_i b_i(\xi) + \sum_i \frac{\tilde{\nu_i}}{\nu_i-\tilde{\nu_i}}
\xi_i b_i(\xi) ]
\frac{\kappa}{v 6 {\pi}^{1/2} \gamma^{3/2}} + \kappa^{2/3}
\ \ .
\end{equation}

It is necessary to express $\nu_i$
through
    $S^{3/2}$.
Then
\begin{equation}
\frac{\tilde{\nu_i}}{\nu_i - \tilde{\nu_i}} \xi_i  =
\frac{v 6 \pi^{1/2} \gamma^{1/2}}{\kappa^{1/3}} \lambda_i
\ \ .
\end{equation}

For the free energy one can approximately get
\begin{equation}
F =  - [\sum_i \xi_i b_i(\xi)]
\frac{\kappa}{v 6 \pi^{1/2} \gamma^{3/2}}
+
\kappa^{2/3} [ 1 - \frac{\sum_i \lambda_i b_i(\xi)}{\gamma} ]
\ \ .
\end{equation}

We have the same analytical structure as the
already observed one after the
shifts
$$
\kappa^{2/3} \rightarrow
\kappa^{2/3} [ 1 - \frac{\sum_i \lambda_i b_i(\xi)}{\gamma} ]
\ \ ,
$$
$$
B \rightarrow
B = \frac{[\sum_i \xi_i b_i(\xi)]}{v 6 \pi^{1/2} \gamma^{3/2}
 [ 1 - \frac{\sum_i \lambda_i b_i(\xi)}{\gamma} ]^{3/2}}
\ \ .
$$

This  model  satisfies  all  necessary  requirements  of  the  self
consistency.

\end{document}